# Optical signature of cascade transitions between moiré interlayer excitons

Qinghai Tan [1,2†], Abdullah Rasmita [1†], Zhaowei Zhang[1], K. S. Novoselov[3], Wei-bo Gao[1, 2*]

[1] *Division of Physics and Applied Physics, School of Physical and Mathematical Sciences, Nanyang Technological University, Singapore 637371, Singapore.*

[2] *The Photonics Institute and Centre for Disruptive Photonic Technologies, Nanyang Technological University, Singapore.*

[3] *Department of Materials Science & Engineering National University of Singapore, 9 Engineering Drive 1, 117575, Singapore.*

†*These authors contribute equally.*

**wbgao@ntu.edu.sg*

**Abstract**

Cascade transition between energy levels has important applications, such as in quantum information protocols and quantum cascade lasers. In two-dimensional heterostructure, the moiré superlattice potential can result in multiple interlayer exciton (IX) energy levels. We demonstrate the cascade transitions between such moiré IXs by performing time- and energy-resolved photoluminescence measurements. We show that the lower-energy moiré IX can be excited to higher-energy ones, facilitating IX population inversion.

## Introduction

Moiré superlattice can emerge when two different two-dimensional (2D) van der Waals crystals are stacked on top of each other (*1-5*). Similar to the superlattice in the multiple quantum well (MQW) system (*6*), the moiré superlattice can result in the creation of the electron minibands (*7-14*). Compared to MQW, the fabrication of 2D heterostructure is simpler since it does not involve epitaxial methods necessary for the former one (*15*). Moreover, the interlayer twist angle can modulate the moiré superlattice potential (*7*), which adds another degree of freedom to the superlattice design.

For optoelectronics applications, the 2D transition metal dichalcogenide (TMD) is one of the most promising 2D materials due to its direct bandgap at the monolayer limit (*16, 17*). The optical emission in this material is dominated by the exciton (i.e., bound electron-hole pair) (*18*). Similar to the electron case, the moiré superlattice in the TMD heterobilayer has been shown to create the minibands for both intralayer (electron and hole in the same layer) (*19-22*), and interlayer (electron and hole in different layers) exciton (*20, 22-25*). Such minibands result in multiple emission and absorption peaks (*19-28*). Regarding the interlayer exciton (IX), its long lifetime and large binding energy are advantageous for realizing high-temperature Bose-Einstein condensate (*29*). Moreover, the confinement provided by the moiré potential can result in the IX quantum emitter array (*26, 30*).

The existence of multiple moiré IX minibands in 2D TMD heterobilayer raises the question of whether there are transitions between these minibands, resulting in a cascade transition (i.e., one transition followed by another one) to the ground state. Cascade transition may result in entangled photon emission (*31, 32*), an important resource in quantum information protocols. Additionally, cascade transitions are essential for quantum cascade lasers (QCL). In the traditional QCL architecture (*33*), the cascade transition between the electron minibands of the MQW superlattice is utilized as the lasing transition. The laser emission energy mainly depends on the superlattice design instead of the active material bandgap. Hence, the QCL design can accommodate a wide range of possible wavelength choices, including the terahertz regime (*34, 35*). A similar scenario may also apply to the moiré IX. Such 2D moiré-based QCL has advantages such as on-chip integration (*36, 37*). It is thus essential to study if the cascade transition and population inversion can be achieved in the moiré IX system.

In this work, we utilized the energy- and time-resolved photoluminescence (PL) measurement to study the optical signature of the transition between multiple moiré IXs. We observed four peaks in the IX PL from the monolayer-$MoS_2$/monolayer-$WSe_2$ sample and studied the time-resolved PL of individual peaks. By analyzing the time-resolved data, we identified the cascade emission between the moiré IX energy levels. Additionally, we show that it is possible to excite the lower-level moiré exciton to a high energy state which decays quickly to the higher-level moiré exciton, effectively creating a 3-level system - the minimum requirement for population inversion.

## Results

Figure 1A shows the optical microscope of one $MoS_2$/$WSe_2$ sample consisting of monolayer $MoS_2$ and $WSe_2$ encapsulated by hexagonal Boron Nitride (hBN). Based on the second harmonic generation measurement, we determine that the heterobilayer are AB-stacked (Fig. S1A in Supplementary Materials). Due to the type II band alignment in this structure, the electrons and holes are separated into the $MoS_2$ and the $WSe_2$ layers, respectively, forming the IX with emission wavelength at around 1200 nm (*38-40*).

Figure 1C shows the IX PL spectrum obtained from the sample under continuous wave (CW) excitation, while Fig. 1D shows the PL under pulse excitation. In both cases, the spectrum exhibits four peaks, indicating the multiple energy levels of moiré IX, agreeable with previous reports (*21-23, 25, 28*). As can be seen from Fig. 1D, these multiple peaks appear regardless of the excitation power. The power dependence of the integrated intensity shows a sub-linear behavior (see Fig. S2A in Supplementary Materials), agreeable with previous reports (*41*). Furthermore, at excitation power $< 1$ μW, the energies of the exciton peaks increase with increasing power (see Fig. S2B in Supplementary Materials) – a signature of dipolar interaction between localized IXs (*42*).

The multiple IX emission can be explained by considering the IX energy level inside the moiré potential trap (*24*) (Fig. 1B). Similar to the case of quantum dots (*43*), multiple bound states can exist for a deep enough potential trap. Considering that the energy difference between the IX is around 30 meV, a total of 4 IX peaks can exist in a trap with a depth of more than 120 meV, agreeable with the theoretical upper bound of 200 meV (see Supplementary Materials Section S1).

### Cascade emission between moiré IX energy levels

The transitions between the moiré interlayer exciton states should result in the time correlation between the emissions from different energy levels. To study this, we performed the energy- and time-resolved PL measurement by exciting the sample with a pulsed laser at 726 nm. The excited $WSe_2$ excitons relax in the femtosecond-to-picosecond time scale to become IX (*44-46*). We then apply optical filters to study the time-resolved

PL from different IX energy levels, shown as the symbols in Fig. 2A for excitation power of 3.88 μW. The emissions from p1 and p2 are obtained using 1175 nm and 1225 nm bandpass filters in the collection arm, respectively, while the combined emission from p3 and p4 is obtained using a 1250 nm longpass filter (see the filtered spectrum in Fig. S3 in Supplementary Materials).

We observed that: **(1)** all IX emissions show finite delay to reach the maximum PL count, **(2)** the delay time of different peaks are different with the delay time of p2 (p3+p4) is similar to the fast decay time of p1 (p2), **(3)** there are two decays components with nanosecond (ns) and microsecond (μs) time scales, respectively (the log-lin plot is plotted in Fig. S4 in Supplementary Materials), and **(4)** different peaks (p1,p2,(p3+p4)) have different fast and slow decay rates. Here, we emphasize that the full width at half maximum (FWHM) of our excitation pulse is less than 500 ps (see the impulse response function in Fig. S5 in Supplementary Materials), while all the observed emission peak delays in Fig. 2A is more than 1.5 ns after the excitation pulse ends. Hence, these delays cannot be attributed to the finite pulse width. Similar phenomena are observed in another sample but with an AA-stacking alignment (Fig. S1B and S6(A-C) in Supplementary Materials).

Figure 2B shows the rate equation model derived from the observations above. It consists of three bright IXs ($B_1$ to $B_3$), three dark ones ($D_1$ to $D_3$), and one high energy state (H) (more discussion on the H state is presented in Supplementary Materials Section S2). All transition rates are power-independent, resulting in multi-exponential decay. We note here that B3 consists of two states corresponding to p3 and p4 transition. However, since the p3 and p4 are detected together in the time-resolved experiment, these two states are treated as one state ($B_3$). Hence, only three bright states are considered in the model instead of four. For a similar reason, only three dark states are considered.

A delayed emission of a bright state (point **(1)**) means that there are transitions from higher energy states to the bright state. The transition from H to $B_1$ (i.e., the highest energy bright IX) is responsible for the delay observed in p1 (i.e., transition from B1 to ground state). Similarly, the delays observed in p2 (i.e., B2-to-ground) are caused by the B1-to-B2 or H-to-B2 transitions. In such cascade transitions, the decrease in the higher-energy emission intensity is accompanied by the increase in the lower-energy emission intensity. This process explains why the delay in low energy emission is the same as the decay time constant of the higher one (point **(2)**). The observed μs decay component (point (**3**)) is due to the long-lived dark states. Three dark states are needed because there are three different slow time constants (point (**4**)).

To verify this model, we fit the model to the experimental data with the transition rates and the initial population (right after excitation pulse ends) of each level as the fitting parameters. The strength of the transition rate varies by several orders and is illustrated with the darkness of the lines in Fig. 2B (see Table S1 in Supplementary Materials for the detailed transition rate values). The fitting results of the time-resolved PL are shown as lines in Fig. 2A, which agrees reasonably well with the experimental data. Fitting results at excitation powers between 0.5 μW to 20 μW also show a good agreement between the model and the data (see Fig. S7 and S8 in Supplementary Materials), supporting the existence of cascade transitions between the moiré IX energy levels. Furthermore, the power independence of the decay rates indicates that the single exciton decay process dominates over the exciton-exciton annihilation (e.g., via Auger recombination (*47*)) within this power range.

**Excited state absorption from moiré IX to a higher energy level**

Remarkably, at high excitation power (e.g., 17.45 μW), instead of the count increase, we see the count rate for (p3+p4) emission drops right after the excitation laser pulse, as shown in Fig. 3A and 3B. Similar drops are also observed for other excitation powers larger than 10 μW (see Fig. S7(E, F) and S8(E, F) in Supplementary Materials). This drop is only possible if the excitation pulse pumps out the residual exciton population from the previous pulse, i.e., the excited state absorption (ESA).

We note that the non-zero count at time = 0 ns is due to the previous pulse. This residual population depends on pulsed laser excitation power and repetition rate. In particular, the residual population is significant if the repetition rate is larger than the emission decay rate. Considering that the slow decay part of (p3+p4) emission is mainly determined by the B3-to-D3 transition rate of ~0.5 MHz (see Table S1 in Supplementary Materials), it is expected that the residual population is significant for a repetition rate of 2.5 MHz used in Fig. 3.

We next discuss possible transitions corresponding to this ESA. Since the pulse excitation energy (~1.71 eV) is much higher than the energy difference between these moiré excitons (~ 30 meV), the ESA is unlikely to happen between the moiré excitons. Considering the energy levels as shown in Fig. 2B, this means that the excitation pulse excites the bright IX ($B_1$-$B_3$) to the H state. To further confirm this ESA process, we note that the ratio between the population of B3 and H state at the end of excitation pulse can be expressed as

$$N_{\mathrm{H}} \big/ N_{\mathrm{B3}} = \alpha_{\mathrm{BH}} P + \alpha_{\mathrm{0H}}\,, \qquad (1)$$

where $N_{\mathrm{H(B3)}}$ is the population of H ($B_3$) state at the end of the excitation pulse, $P$ is the average excitation power, and $\alpha_{\mathrm{BH(0H)}}$ is proportional to the $B_3$ (ground)-to-H absorption rate (see more details in Supplementary Materials Section S3). This ratio is plotted in Fig. 3C. The ratio shows a linear dependence on excitation power, which is agreeable with Eq. (1), further confirming the optically excited $B_3$-to-H transition.

The observed cascade transition and ESA may have a relevant application for THz lasing. Based on the fitting results (Fig. 2B and Table S1 in Supplementary Materials), we note that the H-to-$B_1$ transition is at least three orders faster than the H-to-$B_3$ transition. Considering that the B3 can be pumped to H, this significant difference in the decay rates means that the exciton pumped from $B_3$ to H will quickly decay to $B_1$. This mechanism provides an efficient population transfer from the lower energy IX to the higher one, enabling IX population inversion. Assuming the cascade transition is radiative, this can result in stimulated emission with 30-60 meV (7-15 THz) energy.

**Temperature dependence of IXs delayed emissions and their lifetime**

Finally, we discuss the effect of temperature on the cascade emission's optical signature (i.e., the delayed emission of IX). The time-resolved PL of (p3+p4) emission with an $\tau_{\mathrm{delay}}$ (i.e., the delay between the PL peak and the pulse stop) decreases with increasing temperature. Eventually, the rising time vanishes at temperatures above 70 K.

The temperature dependence can be explained by considering that fast and slow decay rates increase with increasing temperature (Fig. 4(C, D)). The emitting state's outgoing population rate increases as these rates increase, reducing its population growth. As a result, the maximum count is reached at an earlier time. The temperature dependence of the decay rate can be understood as the result of the increase in phonon-assisted exciton decay rate with increasing temperature. As shown in Fig. 4(C, D), the data agree well with the theoretical model taking into account the phonon-assisted exciton decay mechanism (see Supplementary Materials Section S4 for more details on this model).

## Discussion

In conclusion, by performing time- and energy-resolved PL measurements, we observe the cascade emission between moiré IXs. Furthermore, our experimental results indicate the possibility of an efficient population transfer from the lower-energy IX to the higher one. Further development of our findings may result in THz QCL based on the IX in 2D TMD heterostructure.

We note that this moiré based QCL differs from the recently proposed van der Waals-based QCL, where the superlattice is based on MQW design (*48*). Additionally, the demonstrated ESA also shows that the bright IX population can be driven to one quantum state. Such capability combined with the existence of multiple dark IX with a microsecond lifetime and dark-bright coupling could be exploited for realizing high-temperature Bose-Einstein condensate (*29*). Finally, when the IX emission can be driven into the single-photon regime, the IX cascade transition can be used to generate entangled photon pairs (*31*). Our findings thus open up new possibilities to realize novel light sources and many-body quantum states based on moiré physics.

## Materials and Methods

### Sample Fabrication

Monolayer $WSe_2$, monolayer $MoS_2$, and hBN samples are mechanically exfoliated from bulk crystals on PDMS stamps. The heterostructure is stacked layer-by-layer using a dry transfer method onto the $SiO_2$/Si substrate with ultralow doping Si, as described elsewhere (*39*). We then annealed the samples under ultrahigh vacuum (around $10^{-7}$ mbar) at 200°C for 3 hours.

### Optical Characterization

#### *CW and time-resolved PL measurement*

A homemade confocal microscope is used to measure the PL spectra under CW excitation and time-resolved PL. The PL spectra were obtained by a spectrometer (Princeton) with a liquid nitrogen-cooled charge-coupled device (CCD) detector. A 50× long-work distance objective lens with a spot size of around 1 μm [numerical aperture (NA) = 0.65] is used. The sample temperature is cooled using a cryostat (Montana Instruments and Attocube).

For the time-resolved experiments, a 726-nm diode laser (source FWHM <80 ps with a maximum repetition rate of 80 MHz) is used for exciting the sample. For energy and time-resolved PL measurement, a 1064 nm longpass filter combined with bandpass (1175 nm or 1225 nm) or longpass filters (1250 nm) is used to cut the signal.

***SHG measurements***

The second harmonic generation (SHG) of monolayer samples and heterostructures are measured at room temperature. A pulse laser from a Ti:Sapphire oscillator (Spectra Physics, Tsunami) with a peak at around 878 nm, a repetition rate of 80 MHz, and a pulse duration of 100 fs was used as the excitation source. The pulse laser power is around 0.6 mW.

## Figures and Tables

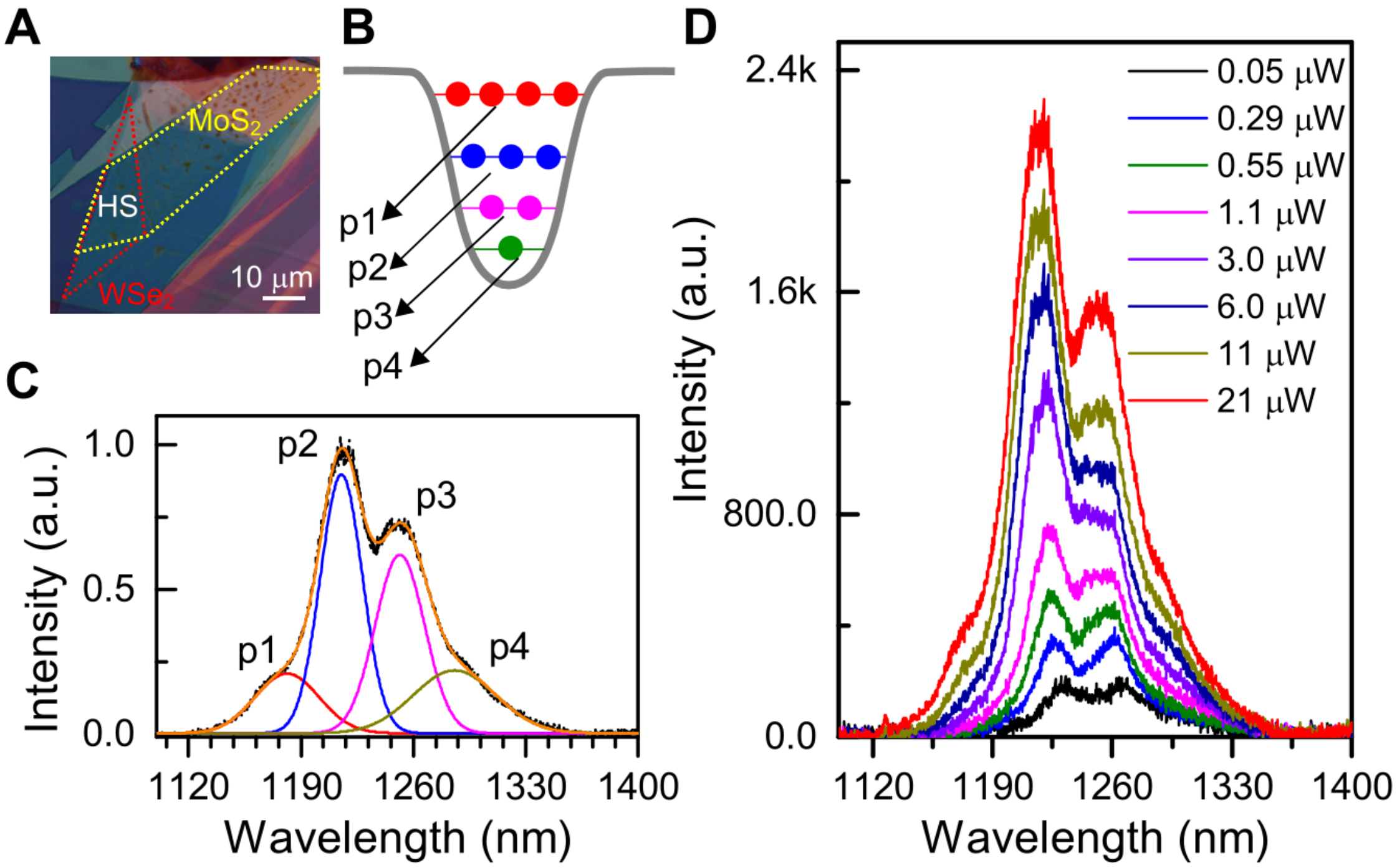

**Fig. 1. Sample and PL spectrum. (A)** Optical microscope image of hBN-encapsulated $MoS_2/WSe_2$ device (Sample 1). **(B)** Illustration of the multiple IX from multiple levels in moiré potential trap. **(C)** IX PL Spectrum at 5 K. A 1064 nm longpass filter is used to obtain the IX PL spectrum (black line). The orange line is the fitting result using four Gaussian peaks (each shown with a different color). **(D)** Power dependence of the PL spectrum under pulse excitation. Regardless of the average excitation power, the PL spectrum always shows multiple peaks.

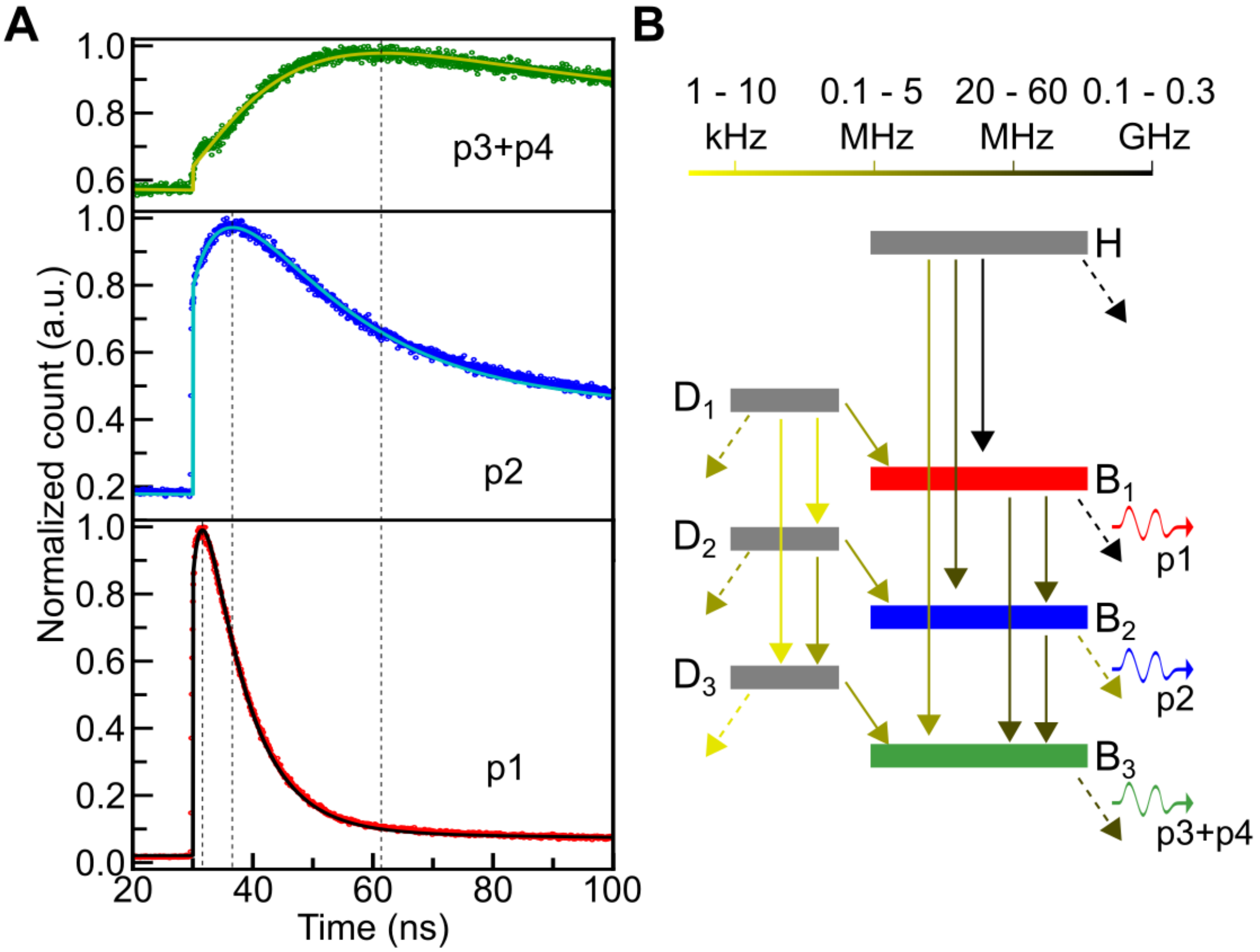

**Fig. 2. Cascade emission between moiré IX. (A)** Energy- and time-resolved IX PL at 4.5 K with average excitation power of 3.88 μW. The symbols are the measured data, while the lines are fitting results using the model in **B**. The p1-p4 is as in Fig.1. The dashed lines indicate the peak position of each emission. The excitation pulse ends at time ~30 ns (see Fig. S5 in Supplementary Materials) **(B)** Rate equation model of moiré IXs transitions. The solid lines indicate transitions between excited states, while the dashed lines are transitions to the ground state. Darker lines indicate faster rates following the color scale. The $D_1(B_1)$-$D_3(B_3)$ are dark (bright) moiré IXs, while H is the high energy state. All transition rate is power-independent.

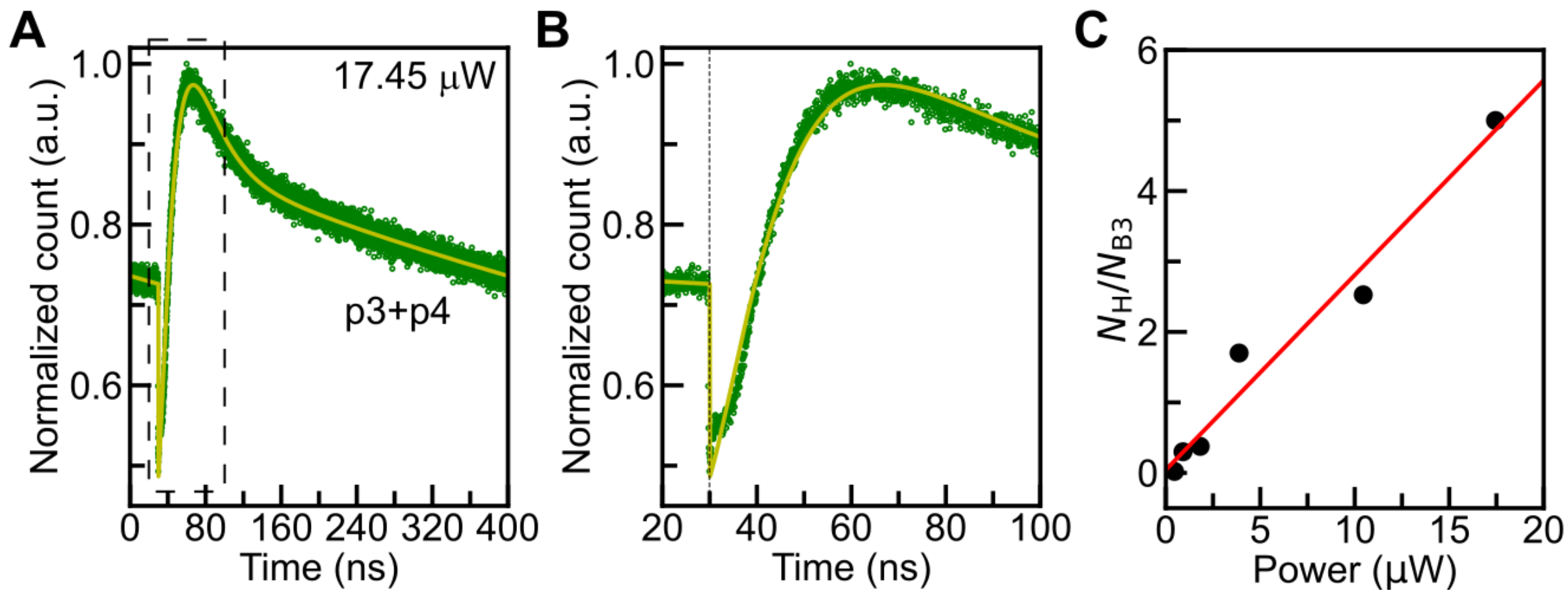

**Fig. 3. Excited state absorption of moiré IX. (A)** Time-resolved IX PL of (p3+p4) emission at 4.5 K with average excitation power of 17.45 μW. The symbols are the measured data, while the lines are fitting results using the model in 2B. **(B)** Zoomed in of the dashed line boxed region in **A**. The dotted line indicates the end of the excitation pulse. The excitation pulse decreases the emission count. **(C)** The ratio between $B_3$ and H population at the initial time (after the end of excitation pulse) vs. excitation power. The symbol is the result obtained from the fitting using the model in Fig. 2B. The error bar is smaller than the symbol size. The line is linear fitting.

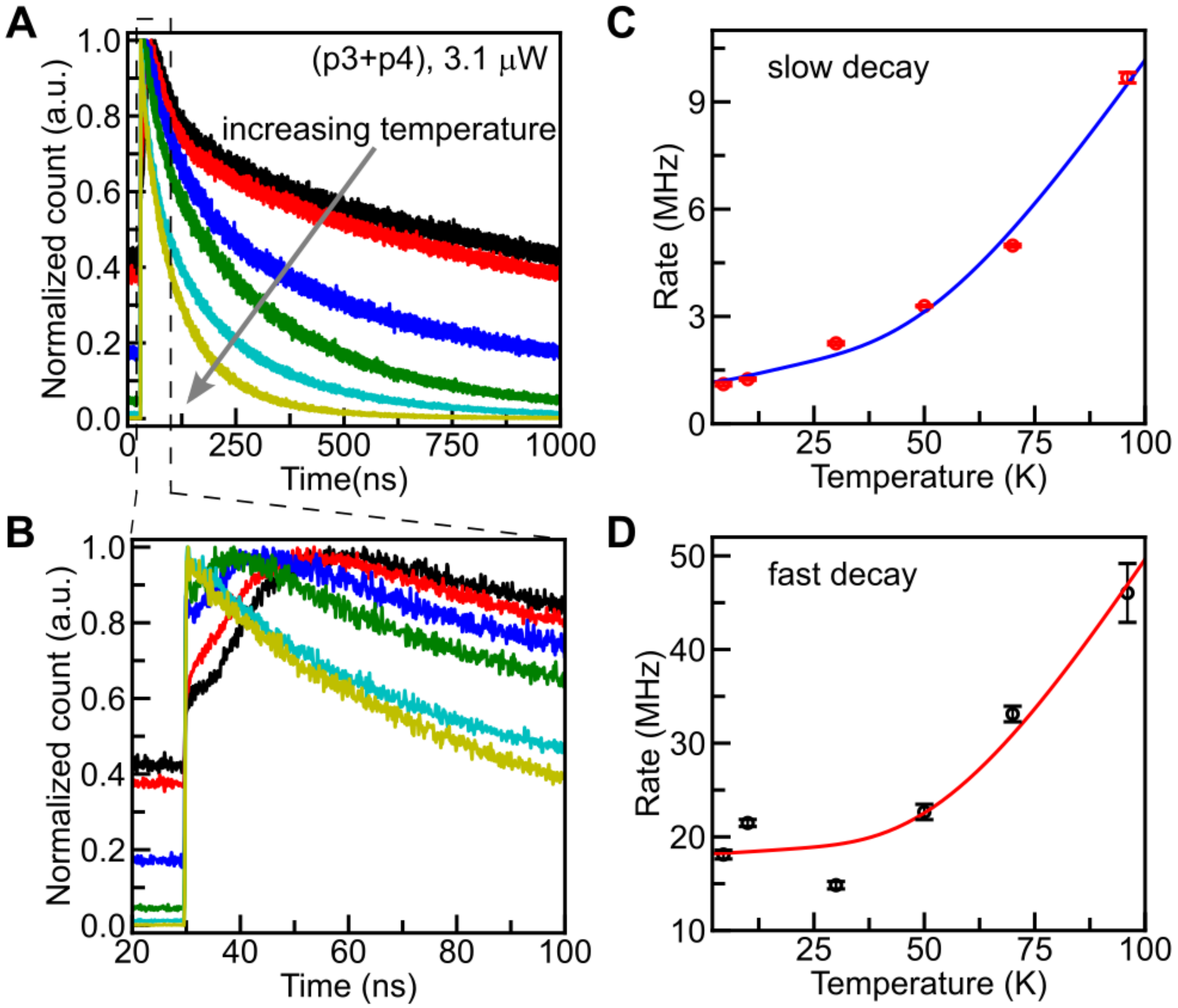

**Fig. 4. Temperature dependence of time-resolved PL. (A)** Time-resolved PL of (p3+p4) emission with an average excitation power of 3.1 μW at various temperatures. The temperatures are 4.5 K, 10 K, 30 K, 50 K, 70 K, and 96 K. **(B)** Zoomed in of **A**. **(C)** Slow and **(D)** fast decay rates. The symbols are obtained from the double exponential fitting of the decaying part (i.e., after the maximum count is reached) of **A**. The error bars represent a 95% confidence interval. The lines are fitting with the theoretical model.

# Supplementary Materials for

## Optical signature of cascade transitions between moiré interlayer excitons

Qinghai Tan, Abdullah Rasmita, Zhaowei Zhang, K. S. Novoselov, Wei-bo Gao*

*wbgao@ntu.edu.sg

### Supplementary Text

S1. Estimation of moiré potential depth

Following the theoretical model in (*23, 24*), the moiré potential depth can be estimated as $V_M = \left(\frac{3a_M V_{pp}}{4\pi\hbar}\right)^2 M_X$, where $V_{pp}$ is the moiré interlayer exciton energy spacing, $M_X$ is the exciton mass which is the summation of electron and hole effective mass, and $a_M$ is the moiré lattice constant. Based on the IX PL spectrum (Fig. 1C in the main text), $V_{pp}$ is around 30 meV. Assuming the effective mass of electron and hole to be $0.5m_e$ (with $m_e$ is the electron bare mass), we got $M_X = m_e$. Assuming that the two layers are perfectly aligned in either AA or AB stacking, the moiré lattice constant can be expressed as $a_M = \frac{a_{Se}a_S}{|a_{Se} - a_S|}$, where $a_{Se(S)}$ is the monolayer $WSe_2$($MoS_2$) lattice constant. The value of $a_{Se}$ is estimated to be in the range of 0.325 to 0.332 nm, while the value of $a_S$ is in the range of 0.312 to 0.319 nm (*49-52*). By using $a_{Se} = 0.325$ nm and $a_S = 0.319$ nm, we obtained $V_M \leq 201$ meV. Experimentally, we found that there are 4 bright IX energy levels with $V_{pp}$ is around 30 meV. This gives an experimental lower bound of $V_M \geq 120$ meV, which is within the theoretical upper bound of $V_M$.

S2. Possible nature of the H state

Here, we discuss the possible nature of the H state used in our model (see Fig. 2B in the main text). Based on Fig. 3C in the main text, we found that $\alpha_{0H} \approx 0$, showing that the pumping from the ground to the H state is negligible. This rules out the possibility that the H state is a zero-momentum exciton, leaving two possibilities for the H state: (1) higher-order interlayer exciton complex, such as interlayer biexciton, or (2) hot exciton.

The first possibility is unlikely because the transition rate from a higher-order exciton complex to exciton usually is faster than the transition rate from exciton to the ground state. In our case, the H-to-$B_3$ transition rate (< 0.2 MHz) is much smaller than the $B_3$-to-ground transition rate (~ 50 MHz), which does not fit the scenario where the H state is a higher-order complex exciton.

Instead, we argue that this H state is a hot exciton state. The transition from IX to hot intralayer exciton state through Auger recombination has been reported before (*53*). Assuming that the H state is the hot exciton state, our experimental results show that the IX can be excited to hot exciton optically. The hot exciton can then decay to become a bright interlayer exciton, effectively creating H-state-assisted ESA transition between the interlayer exciton.

S3. Analysis of simplified 3-level model

In this section, we discuss a simplified 3-level which gives a qualitatively similar exciton dynamic as a 7-level model (Fig. 2B in the main text). The purpose of doing so is to can understand the process involved better. The transition rates involving the 3-level model are illustrated as in Fig. S10A. Here, level 0 is the ground state, while levels 1 and 2 are two excited states.

Our objective is to use the 3-level model to analyze the ESA case (Fig. 3C in the main text), i.e., to get the analytical expression of the ratio between the H state and B state just after the excitation pulse ends, i.e., at point B in Fig. S10B. To do this, we assign the $B_3$ state as level 1 state and the H state as level 2 state. Our strategy is to solve the rate equation for two limiting cases: long pulse duration and short pulse duration cases. Based on the solutions of these two cases, we make a qualitative statement that applies to all cases.

For a long pulse duration, the population at point B does not depend on the population due to the previous pulse. Instead, it can just be obtained using the steady-state analysis of the rate equation. For a 3-level system, we got the following steady-state solution:

$$N_0 = K\left(\gamma_{10}\gamma_{20} + \gamma_{10}\gamma_{21} + \gamma_{12}\gamma_{20}\right), \tag{S1}$$

$$N_1 = K\left(\gamma_{01}\gamma_{20} + \gamma_{01}\gamma_{21} + \gamma_{02}\gamma_{21}\right), \tag{S2}$$

$$N_2 = K\left(\gamma_{01}\gamma_{12} + \gamma_{02}\gamma_{12} + \gamma_{02}\gamma_{10}\right), \tag{S3}$$

where $N_i$ is the population of level *i*, $\gamma_{ij}$ is the transition rate from level *i* to level *j*, i.e., $\frac{dN_i}{dt} = \sum_j \gamma_{ji} N_j - \gamma_{ij} N_i$ , and $K$ is the normalization factor. Since the pumping rate (red lines in Fig. S10) is proportional to excitation power, $P$, we can express $\gamma_{ij}, i < j$ as $\gamma_{ij} = \alpha_{ij} P$. Substituting this to Eq. S1-S3, we obtain

$$N_0 = K\left(\alpha_{12}\gamma_{20}P + \gamma_{10}\left(\gamma_{20} + \gamma_{21}\right)\right), \tag{S4}$$

$$N_1 = K\left(\left(\alpha_{01} + \alpha_{02}\right)\gamma_{21} + \alpha_{01}\gamma_{20}\right)P, \tag{S5}$$

$$N_2 = K\left(\left(\alpha_{01} + \alpha_{02}\right)\alpha_{12}P^2 + \alpha_{02}\gamma_{10}P\right), \tag{S6}$$

Assigning the $B_3$ state as level 1 state and the H state as level 2 state, we then obtain

$$\frac{N_{\mathrm{H}}}{N_{\mathrm{B3}}} = \frac{\left(\alpha_{01} + \alpha_{02}\right)\alpha_{12}}{\left(\left(\alpha_{01} + \alpha_{02}\right)\gamma_{21} + \alpha_{01}\gamma_{20}\right)}P + \frac{\alpha_{02}\gamma_{10}}{\left(\left(\alpha_{01} + \alpha_{02}\right)\gamma_{21} + \alpha_{01}\gamma_{20}\right)} = \alpha_{\mathrm{BH}}P + \alpha_{0\mathrm{H}}, \tag{S7}$$

where $\alpha_{\mathrm{BH}}$ is proportional to $\alpha_{12}$ and $\alpha_{0\mathrm{H}}$ proportional to $\alpha_{02}$, agreeable with Eq. (1) in the main text.

For the short pulse duration case, we first outline the solution for the general case and then take the Taylor expansion up to second order in pulse duration. Referring to Fig. S10B, the time duration can be divided into two stages: pulse-on (stage I) and pulse-off (stage II), see Fig. S10B. The rate equation at these two stages can be written as

$$\frac{d}{dt}\mathrm{N}(t) = \mathrm{M}_i \mathrm{N}(t), \tag{S8}$$

where $\mathrm{N}(t) = [N_0(t)\ N_1(t)\ N_2(t)]$ and $\mathrm{M}_i$ is the rate equation matrix at stage $i$ depending on time. Referring to Fig. S10B and solving (S8) for each stage, we obtain

$$\mathrm{N_A} = \mathrm{Q_I v_I}, \tag{S9}$$

$$\mathrm{N_B} = \mathrm{Q_I E_I v_I} = \mathrm{Q_{II} v_{II}}, \tag{S10}$$

$$\mathrm{N_{A'}} = \mathrm{Q_{II} E_{II} v_{II}}, \tag{S11}$$

where $\mathrm{N}_k$ is $\mathrm{N}(t)$ at point $k \in \{\mathrm{A,B,A'}\}$, $\mathrm{M}_i\mathrm{Q}_i = \mathrm{D}_i\mathrm{Q}_i$ with $\mathrm{D}_i$ is a diagonal matrix, $\mathrm{E}_i = \exp(\mathrm{D}_i t_i)$, and $i$ represents the stage. After several pulses, a quasi-steady-state condition is achieved where $\mathrm{N_A} = \mathrm{N_A}'$. Combined with the fact that the total population is always constant (here, we take it equal to 1), we obtain the following matrix equation

$$\begin{bmatrix} Q_{\mathrm{I}} & -Q_{\mathrm{II}}E_{\mathrm{II}} \\ -Q_{\mathrm{I}}E_{\mathrm{I}} & Q_{\mathrm{II}} \\ \sum_{row} Q_{\mathrm{I}} & \sum_{row} Q_{\mathrm{II}} \end{bmatrix} \begin{bmatrix} \mathrm{v_I} \\ \mathrm{v_{II}} \end{bmatrix} = \begin{bmatrix} 0 \\ 0 \\ 2 \end{bmatrix}, \tag{S12}$$

where $\sum_{row} X$ indicate summation over the rows of $X$. The population at point B can then be obtained from (S12) and (S10). Taking the Taylor expansion up to second order in $t_{\mathrm{I}}$, we obtain

$$\frac{N_{\mathrm{H}}}{N_{\mathrm{B3}}} = \alpha_{\mathrm{BH}} P + \alpha_{0\mathrm{H}}, \tag{S13}$$

where $\alpha_{\mathrm{BH}}$ is proportional to $\alpha_{12}$ and $\alpha_{0\mathrm{H}}$ proportional to $\alpha_{02}$, agreeable with Eq. (1) in the main text. Hence, we obtain Eq. (1) in the main text in both long and short pulse duration cases. Based on this, we infer that Eq. (1) in the main text holds for any pulse duration.

## S4. Modeling of the temperature dependence of the decay rate

The general effect of increasing the crystal temperature is increasing the phonon number. We consider two mechanisms where the phonon-exciton interaction can affect the PL decay rate:

1. the phonon-assisted dark-bright mixing, and
2. the phonon-assisted de-trapping of moiré exciton

Considering these mechanisms, the temperature-dependent rate equation can be modeled as in Fig. S11. The fast and slow decay rates are obtained by solving the corresponding eigenvalue equation.

The fitting result (Fig. 4(C, D) in the main text) is obtained using 0.86 meV (equivalent to 10 times Boltzmann constant ($k_{\mathrm{B}}$)) as the dark-bright splitting and 19 meV as the dominant phonon energy. These values agree with the previous reports that the conduction band splitting in $MoS_2$ is in the order of meV (*54-56*), and the vibrational density of state in $MoS_2/WSe_2$ reaches the maximum value at phonon energy around 19 to 50 meV (*57*).

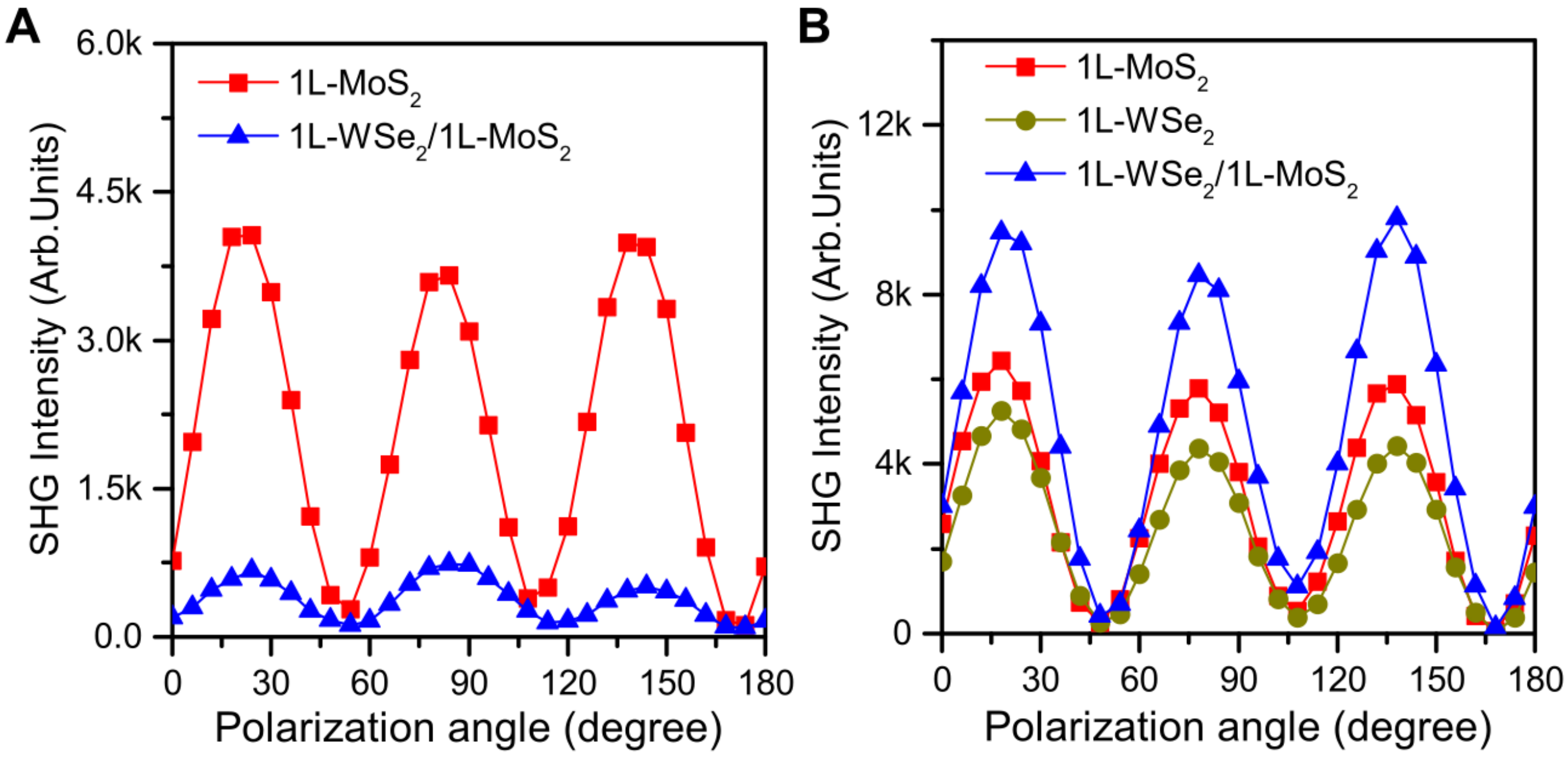

**Fig. S1. SHG measurement of $MoS_2$/$WSe_2$ samples. (A)** Sample 1. **(B)** Sample 2. The SHG measurement results for the monolayer $MoS_2$, $WSe_2$, and heterostructure regions are shown by the red squares, green dots, and blue triangles. For Sample 1, the heterostructure's SHG signal is weaker than those from the monolayers, indicating AB-stacking alignment. For Sample 2, the heterostructure's SHG signal is stronger than those from the monolayers, indicating AA-stacking alignment.

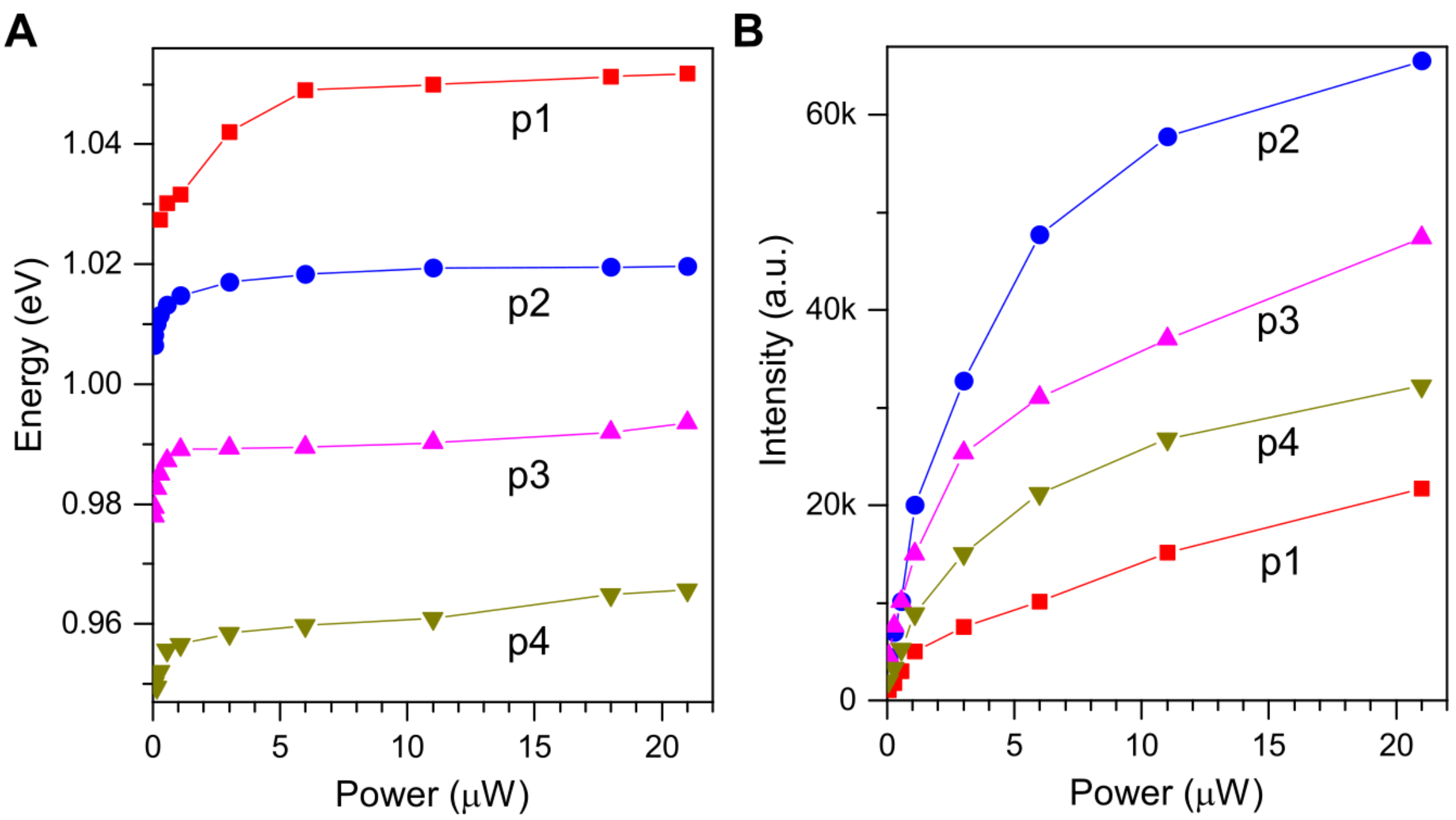

**Fig. S2. Power dependence of the PL spectrum. (A)** Peak position and **(B)** integrated intensity of each moiré IX PL at different power (Fig. 1D in the main text). The error bar for the peak position is less than ±5 meV.

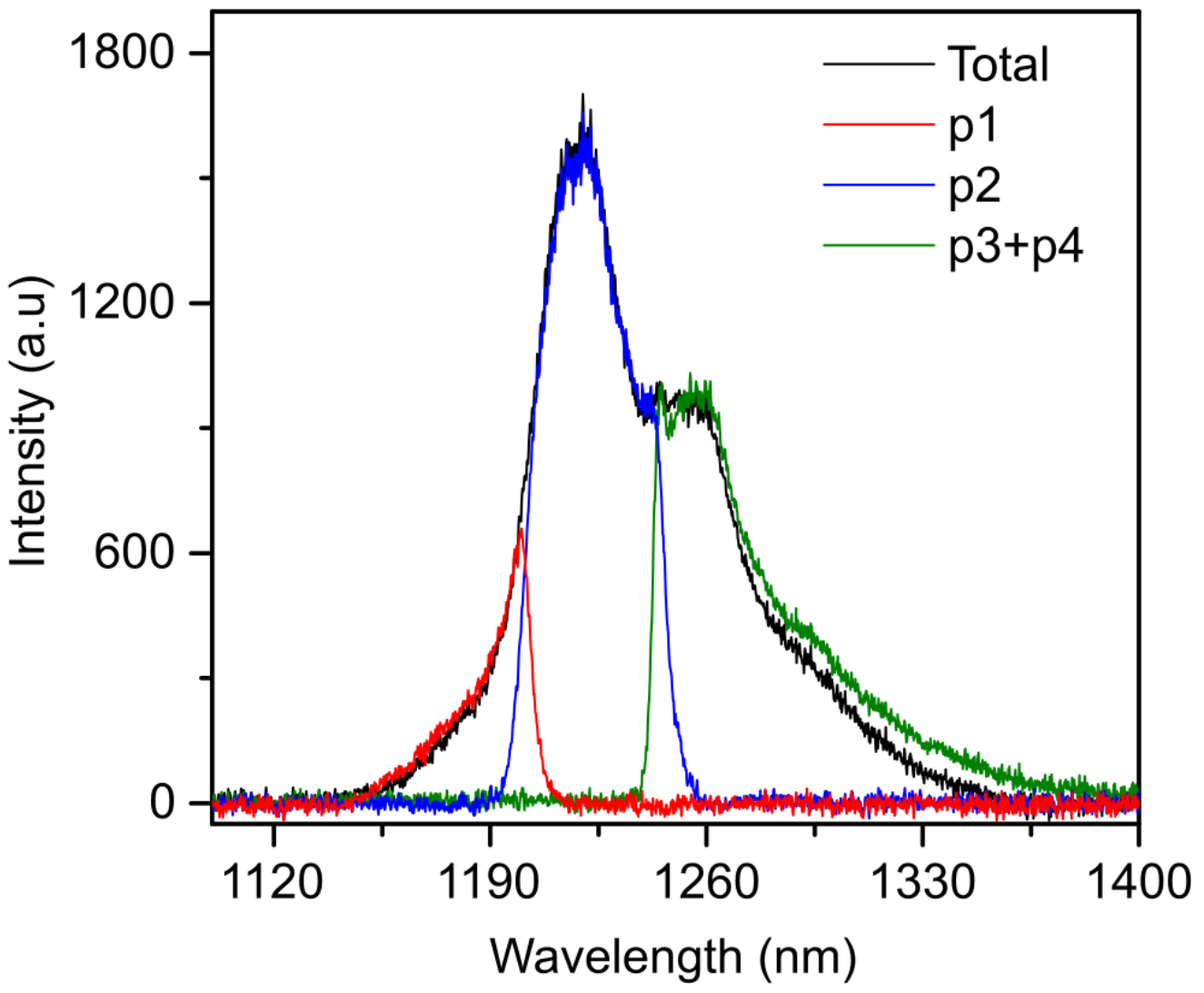

**Fig. S3. Filtered spectrum of sample 1.** The black line is the total IX spectrum obtained by using a 1064 nm longpass filter. The red, blue, and green ones correspond to p1 (1175 nm bandpass), p2 (1225 nm bandpass), and p3+p4 (1250 nm longpass) emissions, respectively.

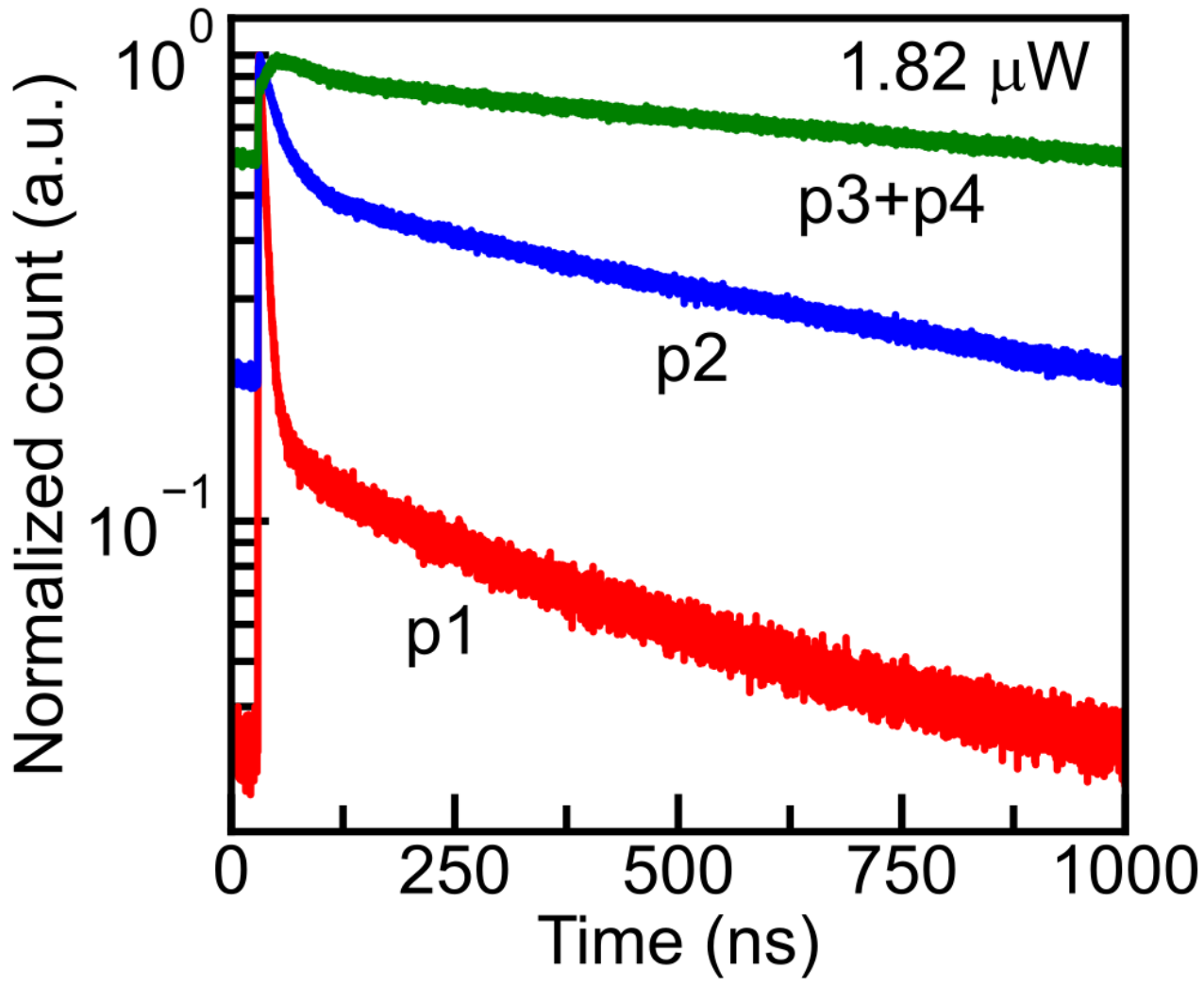

**Fig. S4. Log-linear plot of energy- and time-resolved IX PL at 4.5 K with average excitation power of 1.82 μW.** There are two decays components with ns and μs time scales for each emission.

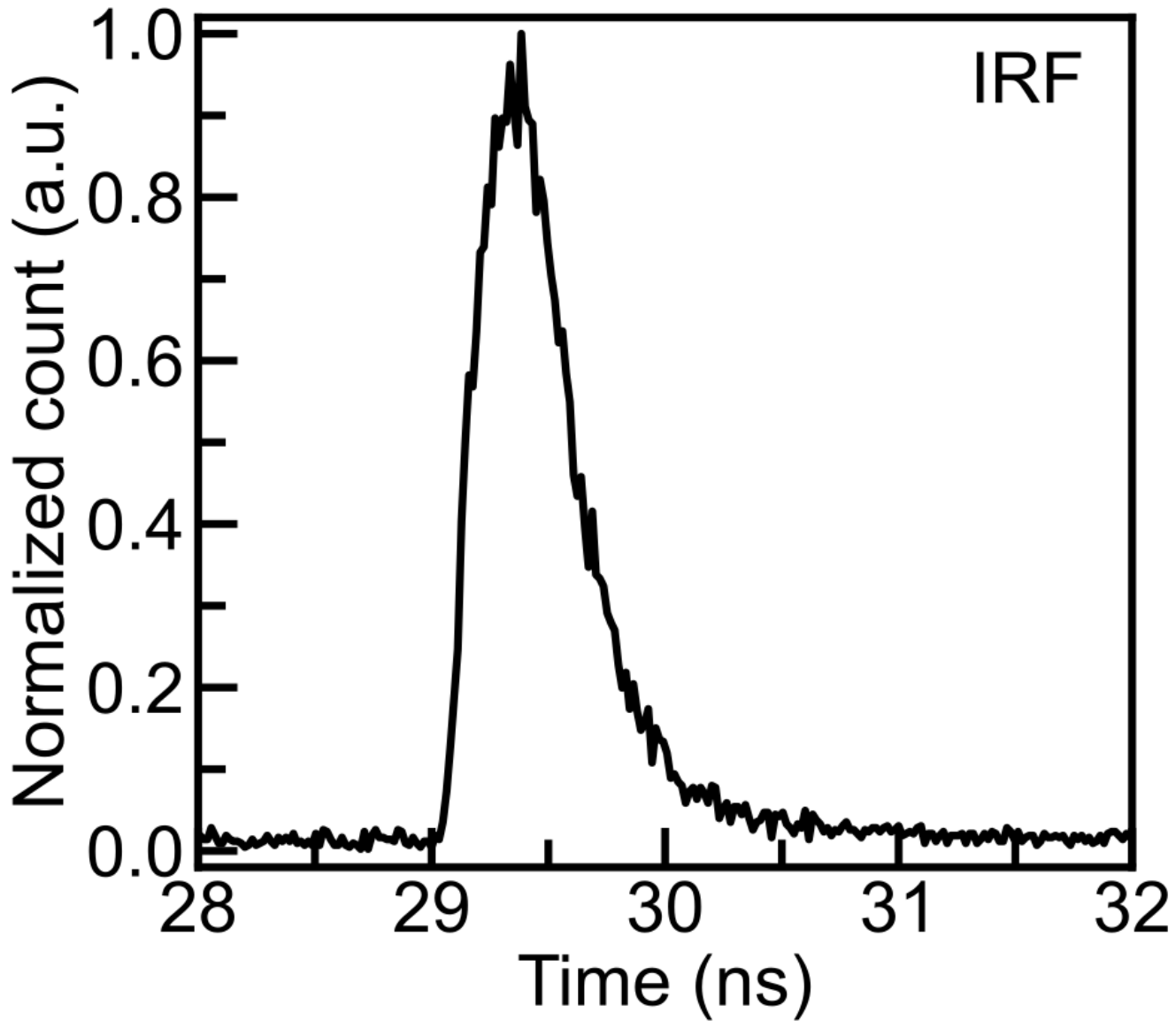

**Fig. S5. Impulse response function of the excitation pulse.** The full width at half maximum (FWHM) is around 470 ps.

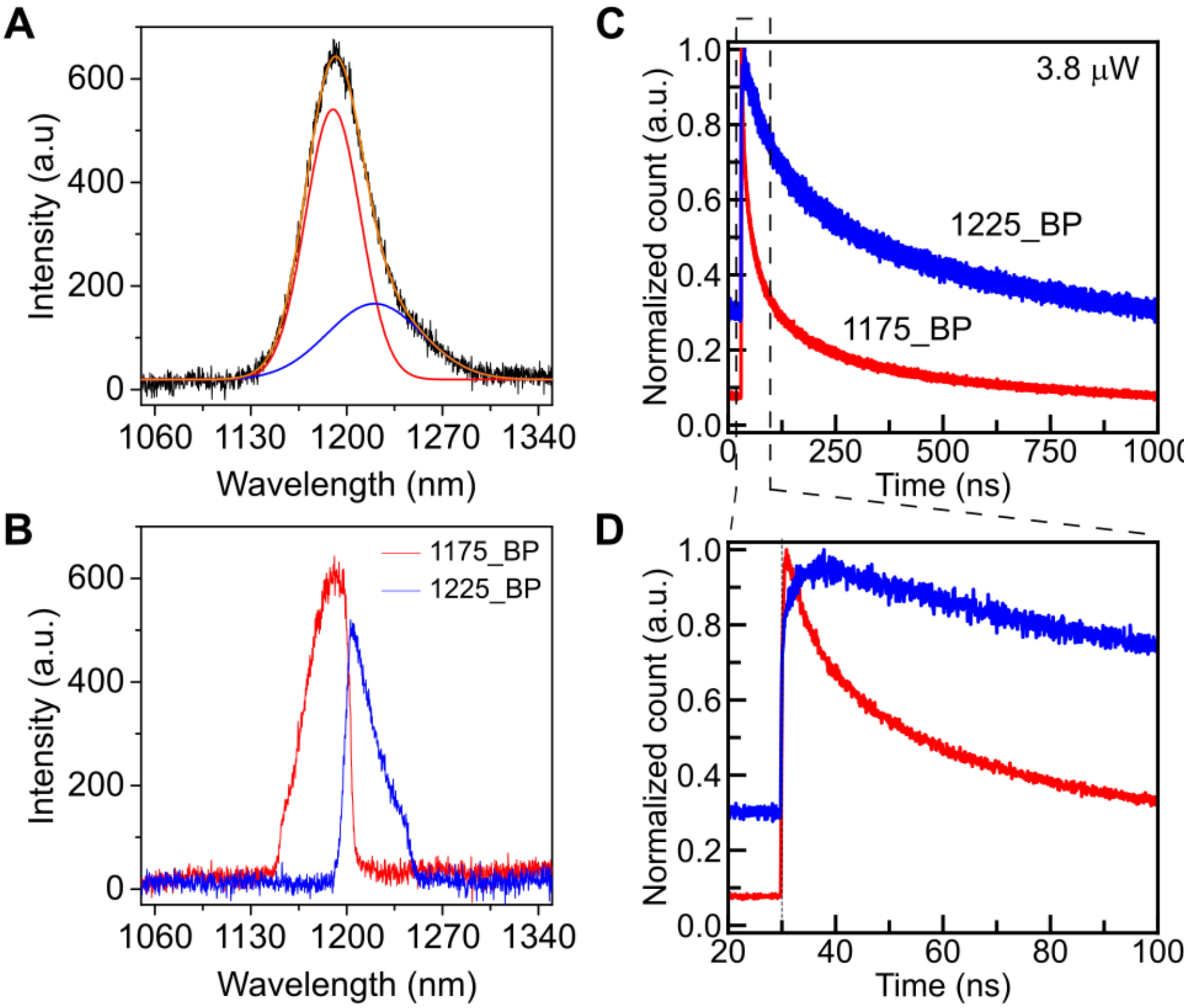

**Fig. S6. Data from another sample (Sample 2). (A)** PL spectrum. A 1064 nm longpass filter is used to obtain the IX PL spectrum (black line). The orange line is the fitting result using four Gaussian peaks (each shown with a different color). **(B)** Filtered PL spectrum. The red and blue lines are the filtered spectrum obtained using 1175 nm and 1225 nm bandpass filters, respectively. **(C, D)** Time- and energy-resolved PL at excitation power 3.8 μW. Similar to the Sample 1 case, the delayed emission is observed.

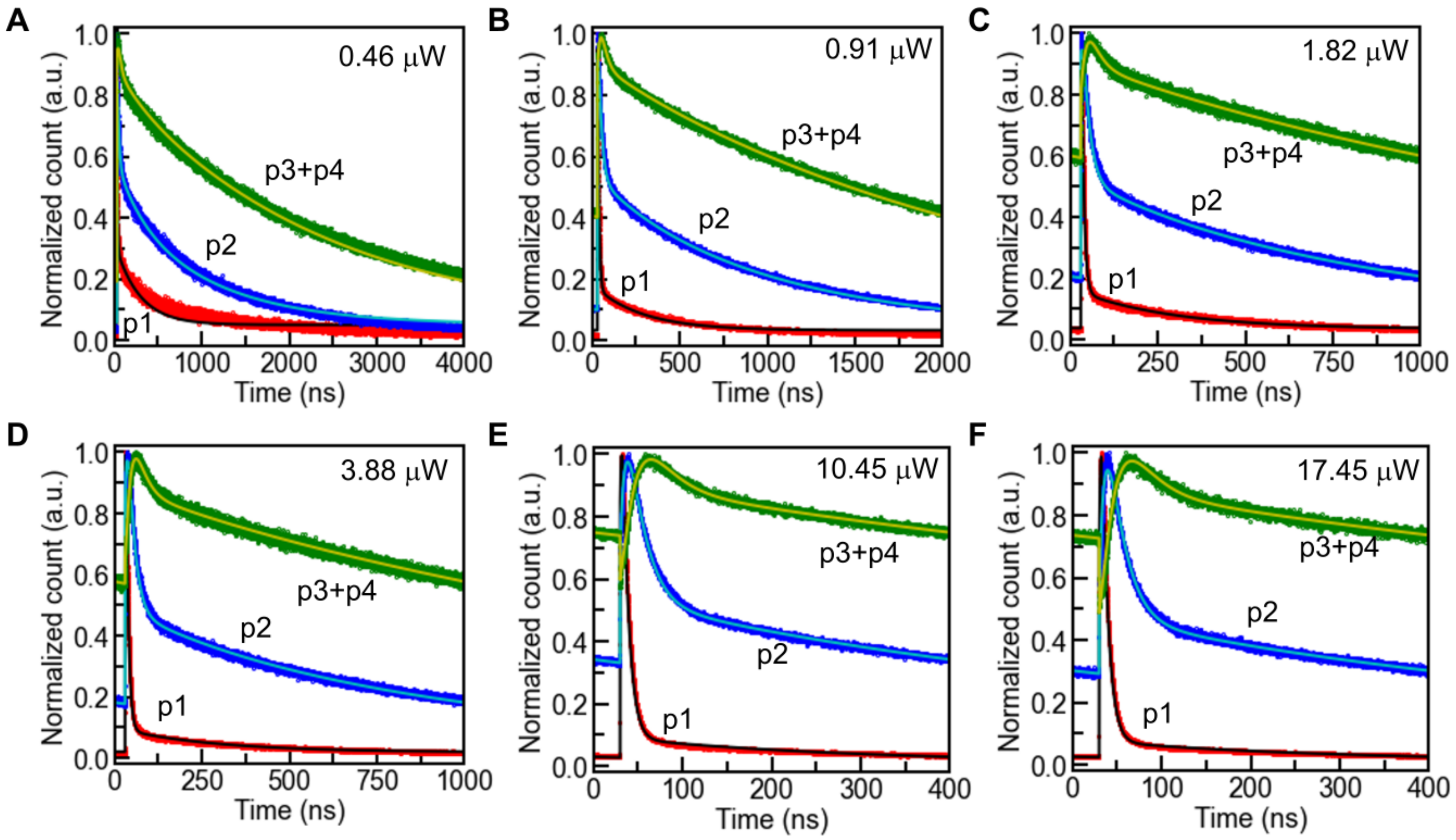

**Fig. S7. Energy- and time-resolved IX PL at 4.5 K at various excitation power.** The symbols are the measured data, while the lines are fitting results using the model described in Fig. 2B in the main text. The p1-p4 is as in Fig.1 in the main text. The fitting for p1 at excitation power less than 0.5 μW is worse than at high power, indicating that the p1 exciton lifetime has power dependence at a low power regime. This power dependence could be attributed to unsaturated exciton-exciton interaction at low power.

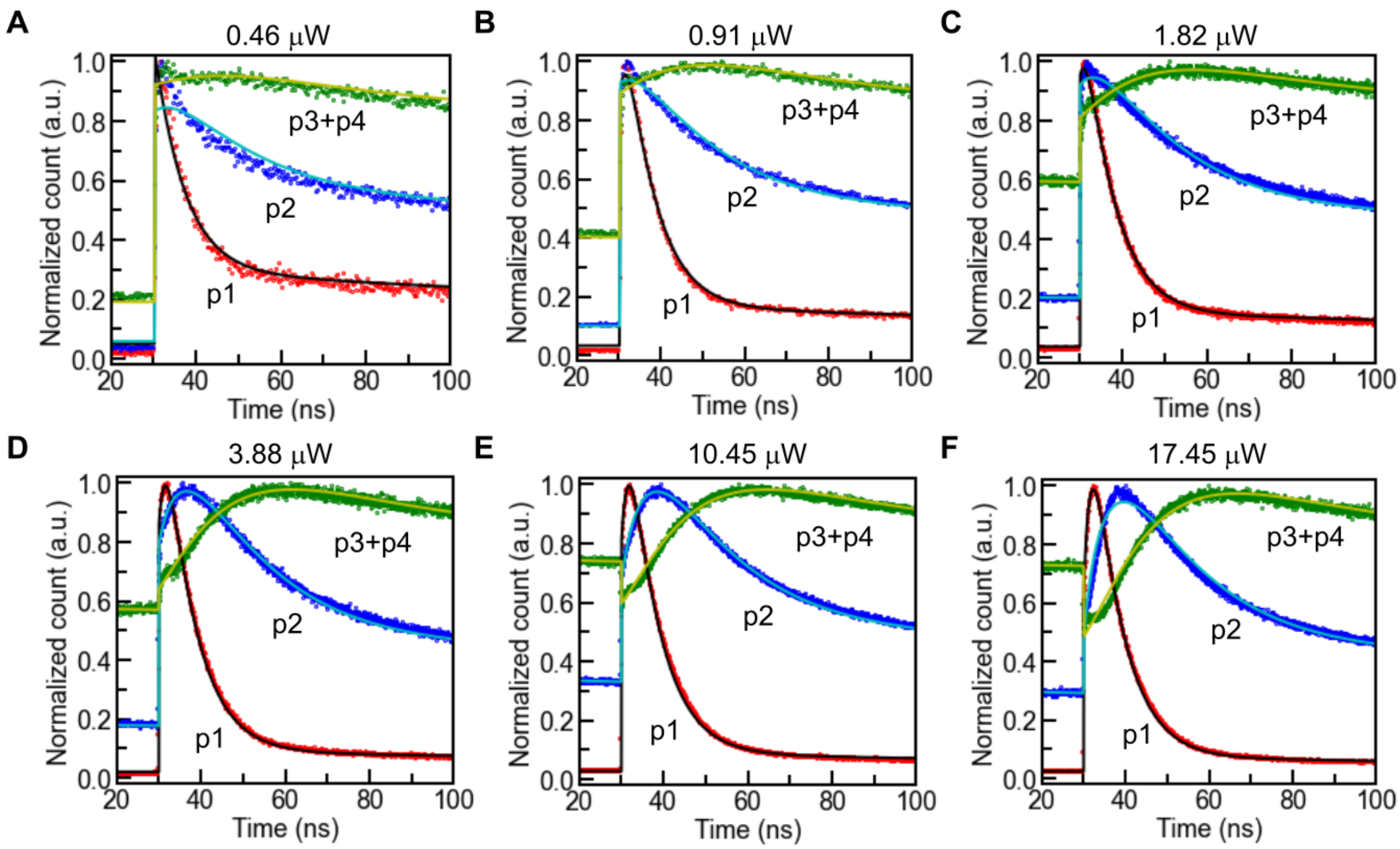

**Fig. S8. Zoomed in of Fig. S7.**

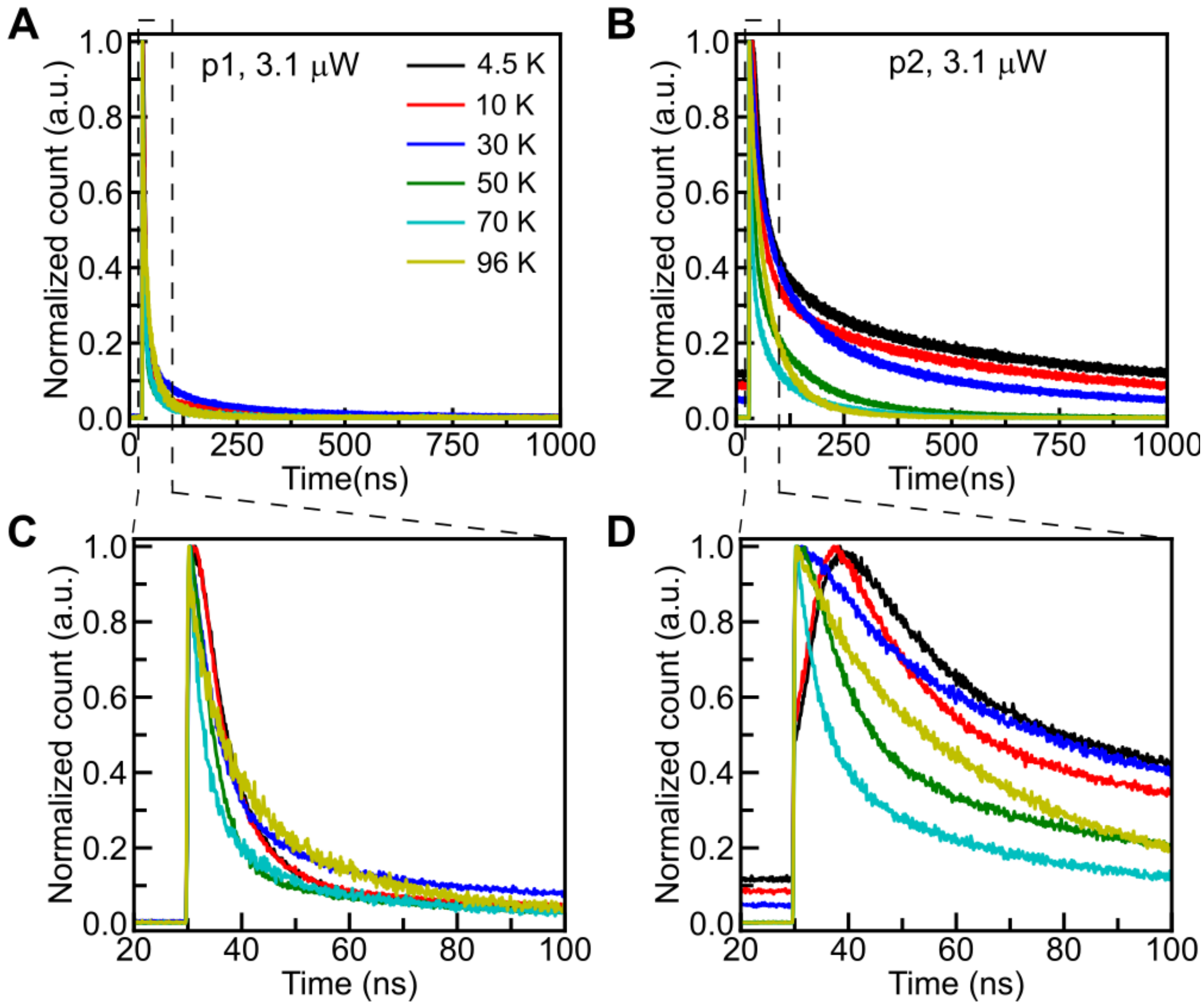

**Fig. S9. Temperature dependence of time-resolved PL.** Time-resolved PL of **(A)** p1 and **(B)** p2 emission with an average excitation power of 3.1 μW at various temperatures. The temperatures are 4.5 K, 10 K, 30 K, 50 K, 70 K, and 96 K. The **(C)** and **(D)** are zoomed in of **A** and **B**, respectively.

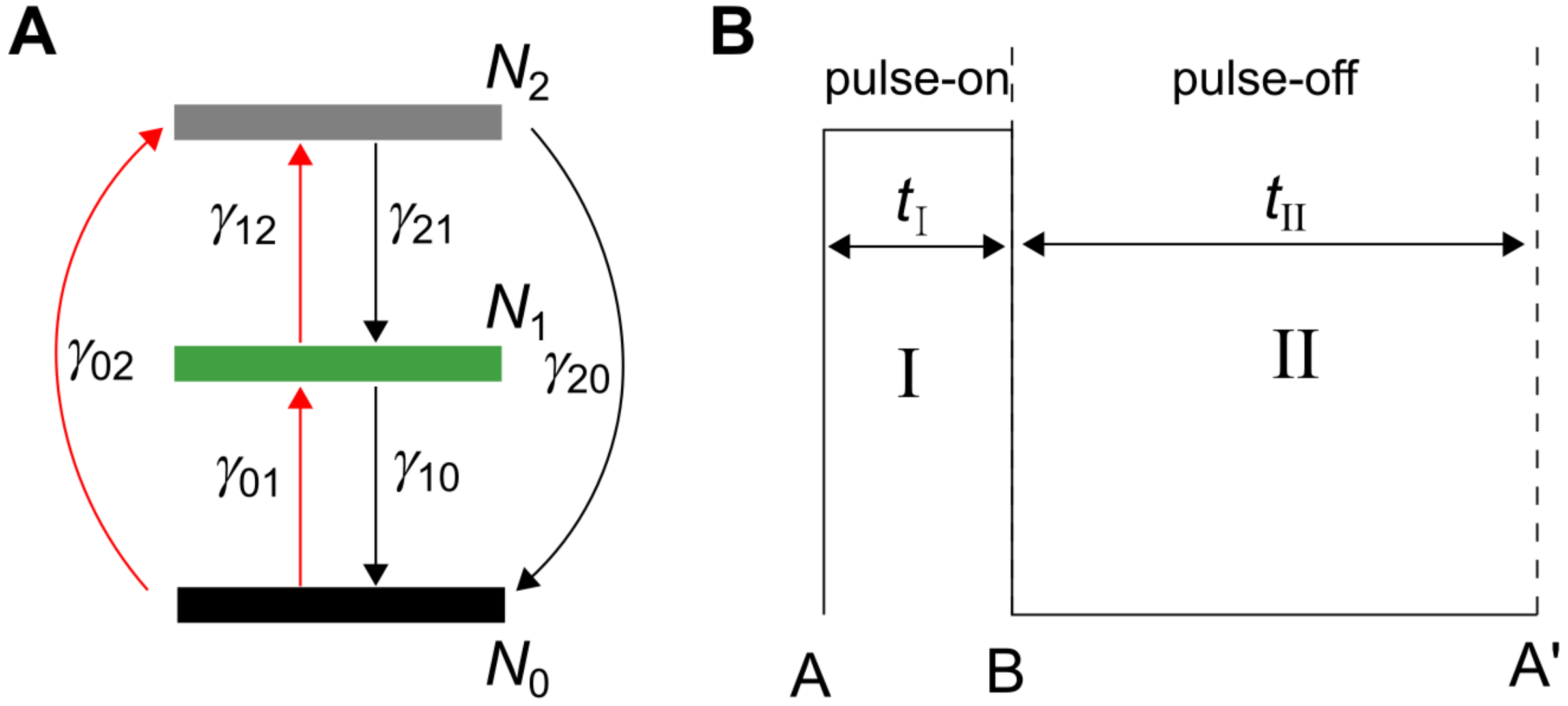

**Fig. S10. Rate equation.** The $N_i$ is the population of level *I*, and $\gamma_{ij}$ is the transition rate from level *i* to level *j*, i.e., $\frac{dN_i}{dt} = \sum_j \gamma_{ji} N_j - \gamma_{ij} N_i$ . (b) Pulse region schematic. The time duration of the pulse-on stage (I) and pulse-off stage (II) are $t_1$ and $t_2$, respectively. The transition rate during one stage is constant, with no transition from lower to higher levels at the pulse-off stage. At the quasi-steady-state condition, the population in A is the same as in A'.

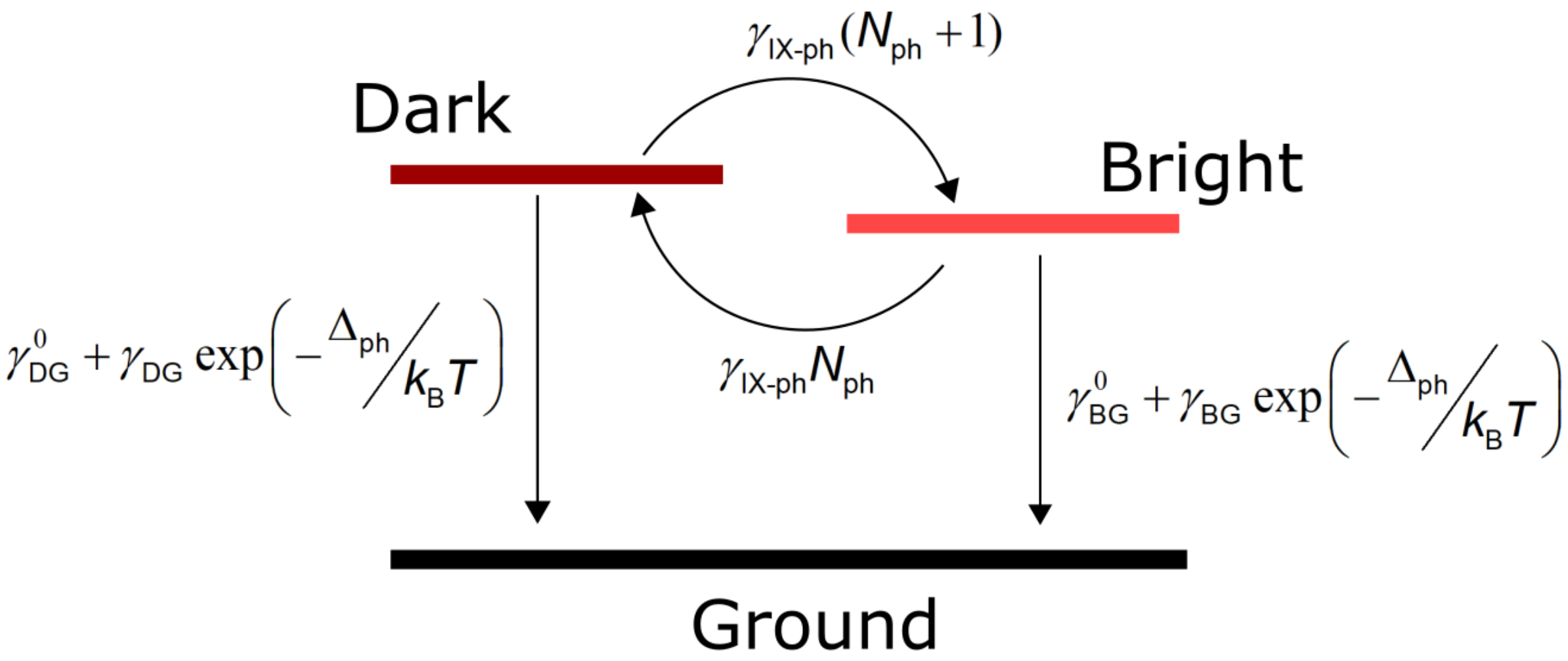

**Fig. S11. Temperature dependence model.** The $\gamma_{\text{IX-ph}}$ is proportional to the phonon-assisted dark-bright exciton mixing, $N_{\text{ph}}$ is phonon number with energy equal to the dark-bright splitting, $\gamma_{\text{BG(DG)}}$ is proportional to the average phonon-bright (dark) exciton coupling, $\Delta_{\text{ph}}$ is the dominant phonon energy, $k_{\text{B}}$ is Boltzmann constant, $T$ is temperature, and $\gamma^0_{\text{BG(DG)}}$ is the bright (dark) exciton decay rate corresponding to mechanisms which do not involve phonon (e.g., radiative recombination and Auger recombination).

**Table S1. Transition rates values based on the fitting result**

| Transition | Value (MHz) |
|---|---|
| $D_1$ to ground | 2.57 ± 0.01 |
| $D_2$ to ground | 0.230 ± 0.003 |
| $D_3$ to ground | 0 ± 0.001 |
| $D_1$ to $D_2$ | 0 ± 0.008 |
| $D_1$ to $D_3$ | 0 ± 0.01 |
| $D_2$ to $D_3$ | 0.419 ± 0.003 |
| $D_1$ to $B_1$ | 1.028 ± 0.002 |
| $D_2$ to $B_2$ | 0.300 ± 0.001 |
| $D_3$ to $B_3$ | 0.438 ± 0.001 |
| H to ground | 221 ± 1 |
| H to $B_1$ | 323 ± 1 |
| H to $B_2$ | 32.7 ± 0.2 |
| H to $B_3$ | 0 ± 0.2 |
| $B_1$ to ground | 59.6 ± 0.1 |
| $B_2$ to ground | 4.4 ± 0. 1 |
| $B_3$ to ground | 49.02 ± 0.09 |
| $B_1$ to $B_2$ | 25.27 ± 0.08 |
| $B_1$ to $B_3$ | 59.6 ± 0.1 |
| $B_2$ to $B_3$ | 51.59 ± 0.08 |